# Coherent wave control in complex media with arbitrary wavefronts

Philipp del Hougne, K. Brahima Yeo, Philippe Besnier, and Matthieu Davy

*Univ Rennes, INSA Rennes, CNRS, IETR - UMR 6164, F-35000, Rennes, France*
Correspondence to philipp.delhougne@gmail.com or matthieu.davy@univ-rennes1.fr.

**Wavefront shaping (WFS) has emerged as powerful tool to control the propagation of diverse wave phenomena (light, sound, microwaves, …) in disordered matter for applications including imaging, communication, energy transfer, micromanipulation, and scattering anomalies. Nonetheless, in practice the necessary coherent control of multiple input channels remains a vexing problem. Here, we overcome this difficulty by doping the disordered medium with programmable meta-atoms in order to adapt it to an imposed arbitrary incoming wavefront. Besides lifting the need for carefully shaped incident wavefronts, our approach also unlocks new opportunities such as sequentially achieving different functionalities with the same arbitrary wavefront. We demonstrate our concept experimentally for electromagnetic waves using programmable metasurfaces in a chaotic cavity, with applications to focusing with the generalized Wigner-Smith operator as well as coherent perfect absorption. We expect our fundamentally new perspective on coherent wave control to facilitate the transition of intricate WFS protocols into real applications for various wave phenomena.**

The interaction of waves with complex scattering matter (multiply-scattering random materials, multimode fibers, chaotic cavities, etc.) results in a complete scrambling of any propagating wavefront, severely hampering many applications in all areas of wave engineering that rely on waves to carry information, for instance, to focus, image or communicate[1]. Nonetheless, by carefully shaping the phase and amplitude profile of a coherent wavefront impinging on a *static* complex medium, these complex scattering effects can to some extent be counteracted (and even harnessed) since they are deterministic[2,3]. We refer to this technique as wavefront shaping (WFS) in the following but this terminology has sometimes also been used to describe various other wave-control approaches in the literature. For the prototypical task of focusing, the shape of the required wavefront can be determined from (highly invasive) field measurements at the target, either via closed-loop iterative schemes[4] or transmission-matrix-based open-loop schemes[5,6], or indirectly with guide-stars that are implanted or virtually created with multi-wave approaches[7–10]. Recently, a new class of open-loop WFS protocols for micromanipulation was introduced by Rotter and co-workers that determines the required wavefront by applying a generalized Wigner-Smith (GWS) operator to the medium's scattering matrix $S$ which must be measured for various perturbations of the target[11,12]. Moreover, WFS-enabled scattering anomalies like coherent perfect absorption (CPA) of incident radiation were recently observed[13,14].

Distinct from WFS are various wave-control efforts based on tuning the scattering properties of a complex medium. On the one hand, this has enabled focusing[15–18] and perfect absorption[19] with single-channel excitation. No coherence of the single-channel wave impinging on the medium can be defined in these cases. On the other hand, complex media have been tuned for "transmission matrix engineering" (TME), i.e. to establish a desired (linear) functional relationship between a coherent input and its associated output wavefront[20–24,18]. TME is "WFS oblivious", i.e., the scattering properties are tuned irrespective of the incident wavefront.[1] Both above-mentioned usages of complex media with tunable scattering properties are therefore incompatible with important goals to date only attainable through WFS, such as micromanipulation[11,12] and CPA[13,14]. Unfortunately, the WFS need for precise individual control in phase and amplitude of the scattering channels to inject the required wavefront $\psi_{WFS}(S)$, as illustrated in Fig. 1(a), thwarts promising applications through costly or impossible hardware requirements.

Here, we overcome this critical hurdle by showing that complex media with tunable scattering properties can be configured *in situ* from $S$ to $S'$ such that a fixed random incident wavefront $\psi_{arb}$ coincides with $\psi_{WFS}(S')$, as illustrated in Fig. 1(b). While WFS adapts $\psi_{WFS}$ to $S$ and the desired functionality, we adapt $S$ to $\psi_{arb}$ and the desired functionality. Thereby, elaborate WFS protocols for micromanipulation or CPA, for instance, can be implemented with an arbitrary wavefront, thus circumventing the vexing need for imposing a specific coherent incident wavefront. Moreover, our approach offers novel functionalities not attainable with WFS. Specifically, a single random wavefront can achieve not only a single but multiple sequential functionalities, which is very attractive for dynamic applications like focusing, micromanipulation or absorption. In addition, counterintuitively, scattering anomalies like CPA can be observed not only with an arbitrary wavefront but also at an arbitrary frequency – in sharp contrast to WFS-based Refs.[13,14].

We experimentally demonstrate our technique for microwaves trapped in a complex scattering enclosure equipped with arrays of reflection-programmable meta-atoms[15,30] that locally reconfigure the boundary conditions[15]. This setting is of direct technological relevance: microwaves used for multichannel wireless communication or sensing often propagate through rich scattering settings like indoor environments, metro stations, airplanes, etc., where such ultrathin programmable metasurfaces are easily added to the walls[31]. First, we apply our scheme to GWS-focusing. Whereas previous WFS-based GWS implementations[11,12] relied on highly invasive manual perturbations of the target, we consider a scenario in which the target naturally induces these variations itself: a backscatter "transmitter"[32] that communicates by modulating the impedance of its port to encode information into ambient waves. We thereby perform a prototypical electronic-warfare counterattack on a spy device such as the infamous Great Seal Bug[33]. Second, we apply our technique to CPA.

---

[1] The inverse design of (usually photonic) wave devices may be seen as a digital predecessor of TME, seeking to fabricate a device with a desired static property[25–29].

Our 3D chaotic cavity shown in Fig. 2(a) is connected to eight channels. Substantial absorption effects on its boundaries imply that $S$ is never unitary in our experiments (see SM). Each of the 304 programmable meta-atoms has two digitalized states corresponding to two opposite electromagnetic responses (see SM). The idea behind their design[34] is to obtain two states emulating Dirichlet or Neumann boundary conditions via a phase difference of roughly $\pi$ at the operating frequency of 5.147 GHz. Since no forward model linking the meta-atom configurations to $S$ exists, we use an iterative optimization procedure (see SM) to identify a configuration for which $\psi_{WFS}(S') = \psi_{arb}$ for a fixed given $\psi_{arb}$.

For any global or local parameter $\alpha$ of the system, the GWS operator can be defined as $Q_\alpha = -iS^{-1}\partial_\alpha S$.[11] The eigenstates of $Q_\alpha$ are invariant with respect to small changes in $\alpha$ so that the outgoing wavefront $\psi_{out}$ remains unchanged apart from a global phase and intensity factor. The associated eigenvalues indicate how strongly the conjugate quantity to $\alpha$ is affected by the scattering process[11]. If $\alpha$ denotes the position of a target, the first or last eigenstate of $Q_\alpha$ are the optimal WFS inputs to focus on or avoid the target, respectively. Micromanipulations other than focusing (e.g. torque or pressure) can be achieved with suitable choices of $\alpha$[12]. The to-date unexplored variable $\alpha$ that we consider in our experiment is the impedance of a scattering port used as backscatter "transmitter". The change of impedance is analogous to the change of dielectric constant in Ref.[12] and therefore the associated conjugate variable can be identified as the integrated intensity of the wave field inside the target[12], in our case the local intensity impacting the targeted port. In other words, injecting the first eigenstate $q_1$ guarantees optimal focusing on our impedance-modulated target. We offer an alternative proof of optimality in the SM.

We now demonstrate that the GWS-enabled effects can be achieved with an arbitrary wavefront $\psi_{arb}$. We define a $7 \times 7$ scattering matrix for our system and connect the eighth port to a switch that can alter its termination (matched load, ML, or open circuit, OC). By approximating $\partial_\alpha S$ with $\Delta S = S_{OC} - S_{ML}$, we estimate $Q_\alpha$, yielding $Q_\alpha \propto -iS_{OC}^{-1}\Delta S$. We reiterate that in our experiment we focus an arbitrary incident wavefront inside a complex medium on a target (the eighth port) without field measurements at the target, without manipulating the target (which auto-modulates its impedance), without knowing the target's location in space ("blind" focusing), and without soliciting the target's cooperation (via a tag or otherwise). That said, we do measure the $7 \times 1$ vector $t$ containing the transmission coefficients between the seven controlled ports and the target, however, solely to experimentally confirm the optimality of the GWS operator. Given $t$, the globally optimal wavefront is easily identified as its phase conjugate $\psi_{pc} = t^*/\|t\|$. Indeed, we observe that the correlation between $q_1$ and $\psi_{pc}$, $C(\psi_{pc}, q_1) = |\psi_{pc}^\dagger q_1|$ ($\psi_{pc}$ and $q_1$ are both normalized), exceeds 0.999 at each iteration of the optimization process (see black dashed line in Fig. 2(b)).

We now judiciously program the meta-atoms so that the eigenvector $q_1'$ coincides with the imposed $\psi_{arb}$ (the prime indicates variables corresponding to the tuned scattering system). We maximize the correlation coefficient between $\psi_{arb}$ and $q_1'$: $C_{GWS} = |C(\psi_{arb}, q_1')|$, where $\psi_{arb}$ is normalized such that $\|\psi_{arb}\| = 1$. The result of an example optimization for a fixed random wavefront is shown in Fig. 2(b) where $C_{GWS}$ reaches 0.9987 after 110 iterations. The ratio between the intensity at the target upon injecting $\psi_{arb}$, $T(S', \psi_{arb}) = |t'^T \psi_{arb}|^2$, and the intensity that would be obtained by focusing with the optimal wavefront $q_1' = \psi_{pc}(S') = t'^*$, $T(S', q_1') = \|t'\|^2$, is equal to $C_{GWS}^2$ and hence converges to unity as the scattering matrix approaches the optimized one $S' \to S_{opt}$ (see Fig. 2(c)). Compared to the average intensity $\langle T(S_{rand}, \psi_{arb})\rangle$ delivered by $\psi_{arb}$ to the targeted port in a random unoptimized system, we achieved with $T(S_{opt}, \psi_{arb})$ an intensity enhancement by a factor of 5.75 in this specific realization.

Of course, the highest achievable intensity $T(S', q_1')$ (black line in Fig. 2(c)) is a statistically distributed quantity such that it fluctuates over the course of the optimization due to the changes of meta-atoms' configurations. A systematic investigation based on 29 realizations with different random $\psi_{arb}$ in Fig. 3(a) reveals that the distributions of $T(S_{rand}, \psi_{pc}(S_{rand}))$ (black) and $T(S_{opt}, \psi_{arb})$ (red) are not identical. While tuning the system's scattering properties enhances the intensity at the targeted port on average by a factor of 4.8, coherent wave control would have enabled an average improvement by a factor equal to the number of incoming channels $M = 7$. We attribute the difference between the two distributions to the presence of an unstirred field component in our system that is not impacted by the meta-atoms, and more specifically to its correlation with $\psi_{arb}^*$. The presence of an unstirred field component is evidenced in Fig. 3(b) in which the clouds of values that different entries of $t$ take for a series of random configurations of the meta-atoms are seen to not be centered on the origin of the Argand diagram. We can thus interpret $t$ as superposition of a stirred component $\Delta t$ and an unstirred component $t_0 = \langle t \rangle$, where $\langle ... \rangle$ denotes averaging over random metasurface configurations: $t = t_0 + \Delta t$. To quantify the relative importance of the two contributions, we introduce the parameter $\kappa = \langle \|\Delta t\|^2 \rangle / \|t_0\|^2$. We estimate $\kappa = 0.18$ for our system. $\kappa \to \infty$ ($\kappa \to 0$) indicates that the programmable meta-atoms offer a perfect (vanishing) degree of control over the wave field.

The degree of correlation $|C(\psi_{arb}, t_0^*/\|t_0\|)|$ determines the performance of our approach benchmarked against coherent wave control, as evidenced with experimental data in Fig. 3(c). To develop a deeper understanding of this dependence, we consider the two extreme cases of unity and zero correlation between $\psi_{arb}$ and $t_0^*$. The goal of the optimization is to tweak $\Delta t$ such that $t'^* = t_0^* + \Delta t'^*$ is collinear to $\psi_{arb}$. If $t_0^*$ is already collinear to $\psi_{arb}$, we only need to make sure that $\Delta t'^*$ is also collinear. The magnitude of $t'$ can therefore be expected to be rather large, yielding large values of $T(S_{opt}, \psi_{arb})$ of the same order as $T(S_{rand}, \psi_{pc})$. In contrast, if $t_0^*$ is perpendicular to $\psi_{arb}$, the stirred field $\Delta t'$ must additionally counterbalance the contribution of $t_0^*$ such that the contribution of $t_0$ to $t'$ is of destructive nature. The magnitude of $t'$ is therefore rather small, resulting in rather small values of $T(S_{opt}, \psi_{arb})$. The achievable value of $T(S_{opt}, \psi_{arb})$ should therefore generally increase with $|C(\psi_{arb}, t_0^*/\|t_0\|)|$, subject to the typical realization-dependent fluctuations in random systems as seen in Fig. 3(c). Fundamentally, this understanding implies that (i) in principle the distributions of $T(S_{rand}, \psi_{pc}(S_{rand}))$ and $T(S_{opt}, \psi_{arb})$ can coincide if sufficient programmable meta-atoms are used, and (ii) in cases where one can determine $t_0$ non-invasively, one can purposefully chose $\psi_{arb}$ to circumvent limitations due to the unstirred field component.

Having demonstrated that our technique enables the use of an arbitrary wavefront to implement GWS-driven coherent wave control for micromanipulation, we now illustrate the versatility of our approach by also applying it to the scattering anomaly of CPA. CPA is a generalization of the critical coupling condition to multi-channel systems in which a zero eigenvalue of the scattering matrix can be accessed by injecting via WFS the corresponding eigenvector $\psi_{CPA}$[35–37]. Then, $\psi_{out} = S\psi_{CPA} = 0$ and all incident radiation is absorbed. The crux of realizing CPA lies in the need to balance excitation and attenuation rate of the system so that $S$ has a zero-valued eigenvalue. This was first achieved with carefully engineered media of typically very regular geometry[38–42]. In static complex media[43,44], both operating frequency and attenuation (which had to be dominated by a single localized loss center) were treated as free parameters, in order to identify a setting for which $S$ had a zero eigenvalue[13,14]. Despite many promising applications in wave filtering, precision sensing and secure communication, these experimental protocols are far too complicated. Recently, the latter constraints were lifted by combining WFS with a tunable complex medium: the scattering properties were optimized such that $S$ had a zero eigenvalue at a desired frequency without explicitly controlling the attenuation level[19,45,46]. The vexing requirement for WFS, however, remained.

We now show that using programmable meta-atoms, $S$ can not only be modified such that it has a zero eigenvalue, but that additionally the corresponding eigenvector coincides with a fixed arbitrary wavefront $\psi_{arb}$. For this set of experiments, we consider the $8 \times 8$ scattering matrix involving all eight ports and the optimization objective is to achieve zero reflection $R' = \|S'\psi_{arb}\|^2$. Obviously, in such a state the incident radiation is not channeled to a single localized loss center because global absorption effects dominate in our cavity; however, this is irrelevant for the aforementioned enticing CPA applications. Given our limited number of programmable meta-atoms, we relax the optimization problem by treating the frequency as a free parameter within a 24 MHz interval around 5.147 GHz. The size of this interval is of the same order as the spectral field-field correlation length (see SM). An example result of the optimization for a fixed arbitrary wavefront is shown in Fig. 4(a). The reflection coefficient reaches a value as low as $R_{CPA} = 1.05 \times 10^{-5}$ ($-49.8$ dB), displaying the extremely narrow dip that is a hallmark feature of CPA. We thereby observe the very special CPA condition in a complex scattering system without having control over the incident wavefront nor over the attenuation in the system. The distribution of $R$ found with 700 random configurations displayed in Fig. 4(b) underlines that CPA is an extremely rare event[43]. Our observed $R_{CPA}$ is four orders of magnitude below the average of $R$. For completeness, we also seek with the same $\psi_{abs}$ a configuration for which $R$ is maximal which corresponds to as little absorption of the incident radiation by the medium as possible. The maximal reflection coefficient is found to be $R_{anti-CPA} = 0.275$ (-5.9 dB) which corresponds to an enhancement by a factor of 3.45 relative to the average of $R$ over random configurations. The homogeneous distribution of attenuation in our system means that while $R_{CPA}$ can reach almost zero and hence enable CPA with a fixed arbitrary wavefront, perfect reflection is impossible[19] and $R_{anti-CPA}$ is always below unity.

In summary, we have put forward an idea for how to implement intricate WFS protocols for complex scattering media without the conventionally required coherent multichannel wave control. We showed that by doping the complex medium with programmable meta-atoms, its scattering matrix can be tweaked such that the necessary wavefront for a desired WFS protocol required for micromanipulation or a scattering anomaly coincides with a fixed arbitrary wavefront. Our proof-of-principle microwave experiments have immediate technological relevance in electronic warfare[33], precision sensing[45], wave filtering and secure wireless communication[19].

Looking forward, an important avenue for future explorations is to identify suitable implementations of our scheme for other wave phenomena, notably light and sound. For light in multimode fibers, current technology already enables built-in liquid-crystal meta-atoms[47] or the external introduction of controlled perturbations with piezoelectric modulators[18]. For biological tissue, we envision the use of magnetic particles[48] or microbubbles[49] that can be wirelessly controlled via external magnetic or acoustic fields, a procedure whose invasiveness would be comparable to the common use of fluorescent markers or contrast agents.

**Figures**

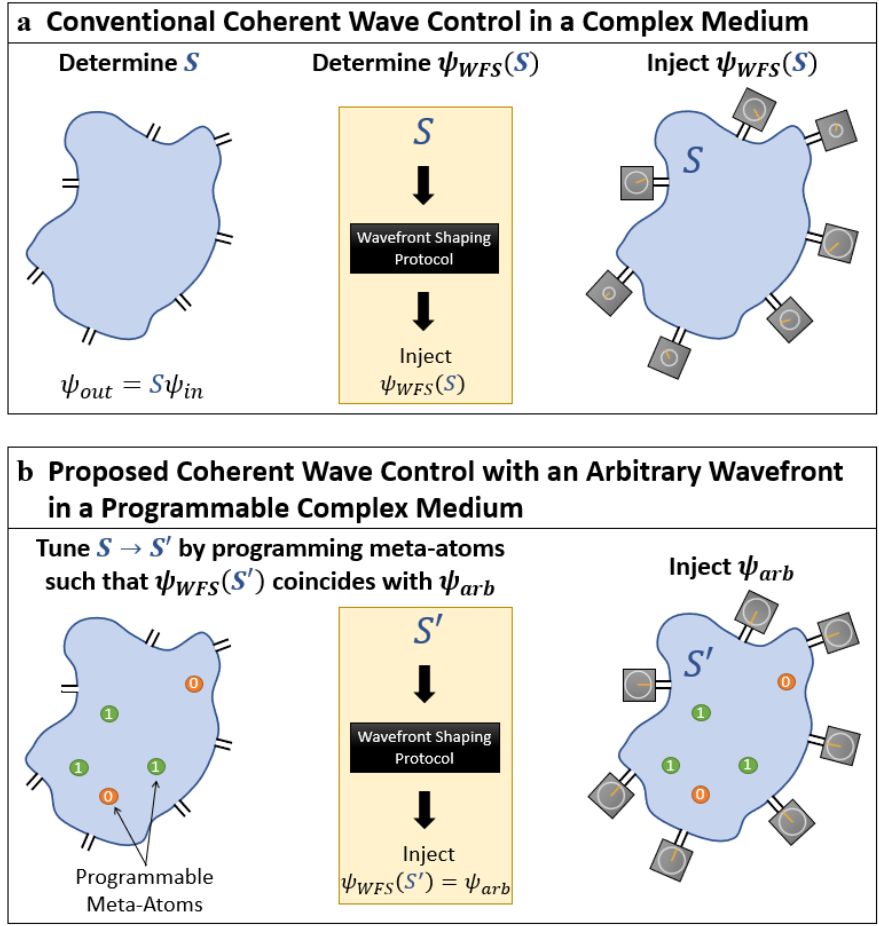

**Fig. 1.** (a) Conventionally, first, the scattering matrix $S$ of a given complex is measured; second, the desired WFS protocol is applied to $S$; third, the obtained $\psi_{\mathrm{WFS}}$ (illustrated as phasors) is injected into the system. Each channel requires independent control of amplitude and phase to inject $\psi_{\mathrm{WFS}}$. (b) In our proposal, the complex medium is doped with programmable meta-atoms, here 1-bit programmable meta-atoms with two possible digitalized states "0" and "1". By judiciously programming the meta-atoms, the system's scattering matrix can be tweaked (from $S$ to $S'$) such that the required wavefront for a desired WFS protocol coincides with a fixed arbitrary wavefront: $\psi_{\mathrm{WFS}}(S') = \psi_{\mathrm{arb}}$.

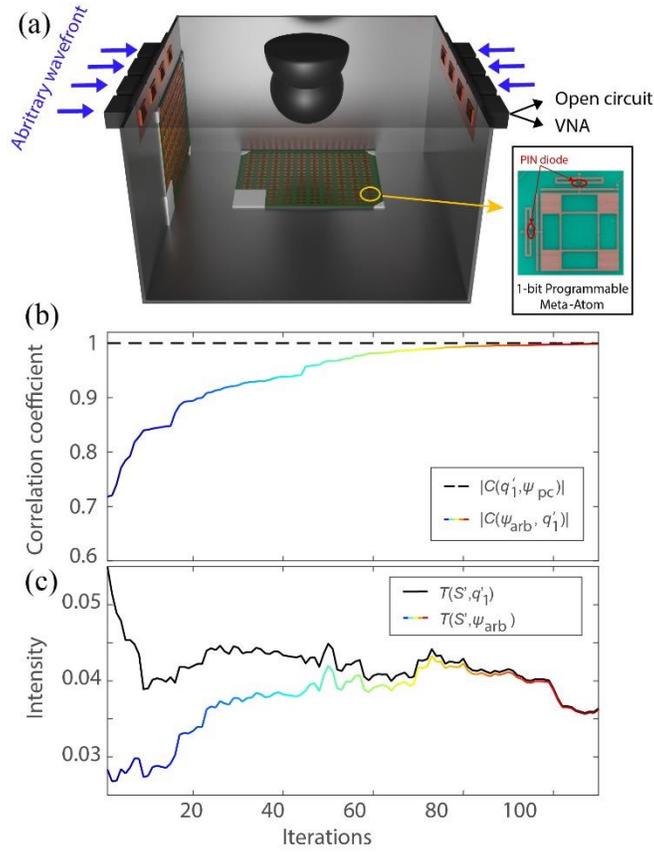

**Fig. 2.** (a) Experimental setup: multi-port irregularly shaped cavity equipped with two arrays of 152 programmable meta-atoms. The inset shows one meta-atom. For our GWS experiments, the eighth port is switched between OC and ML terminations. (b) Example iterative optimization maximizing $\mathcal{C}_{GWS} = |q'_1 \psi^\dagger_{\mathrm{arb}}|$ (colored). Throughout the optimization, $C = |q'_1 \psi_{\mathrm{pc}}(S')|$ is very close to unity (black-dashed line). (c) Variations of intensity $T(S', \psi_{\mathrm{arb}})$ (colored) and optimal value $T(S', q'_1)$ (black) over the course of the optimization.

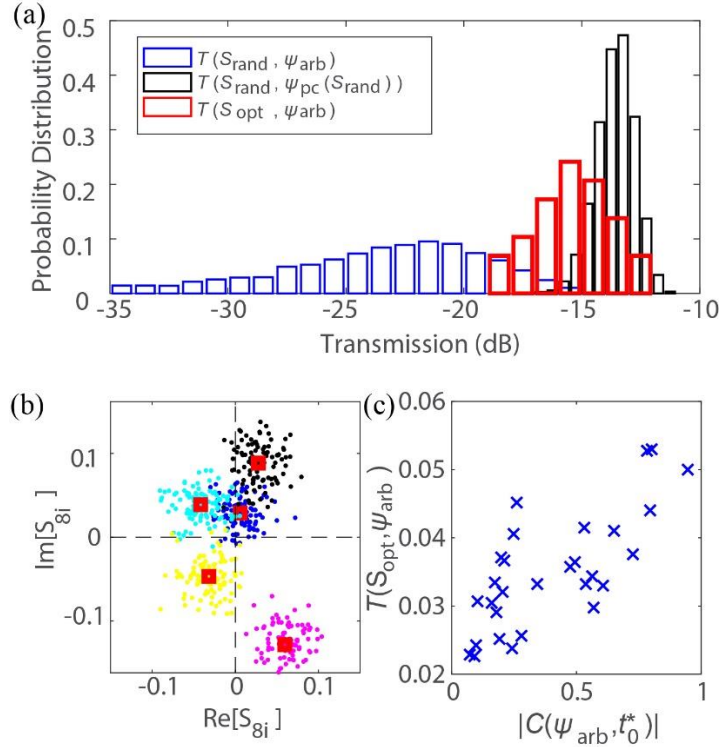

**Fig. 3.** (a) Distribution of intensity transmitted to the target on a logarithmic scale for an arbitrary impinging wavefront on a random system ($T(S_{\mathrm{rand}}, \psi_{\mathrm{arb}})$, blue), the optimal wavefront impinging on a random system ($T(S_{\mathrm{rand}}, \psi_{\mathrm{pc}}(S_{\mathrm{rand}}))$, black), and an arbitrary wavefront impinging on a system optimized for that arbitrary wavefront ($T(S_{\mathrm{opt}}, \psi_{\mathrm{arb}})$, red). (b) Visualization in the complex plane of the transmission coefficients from five ports to the targeted port for 100 random configurations of the meta-atoms. (c) Dependence of $T(S_{\mathrm{opt}}, \psi_{\mathrm{arb}})$ on the degree of correlation between $\psi_{\mathrm{arb}}$ and $t_0^*$.

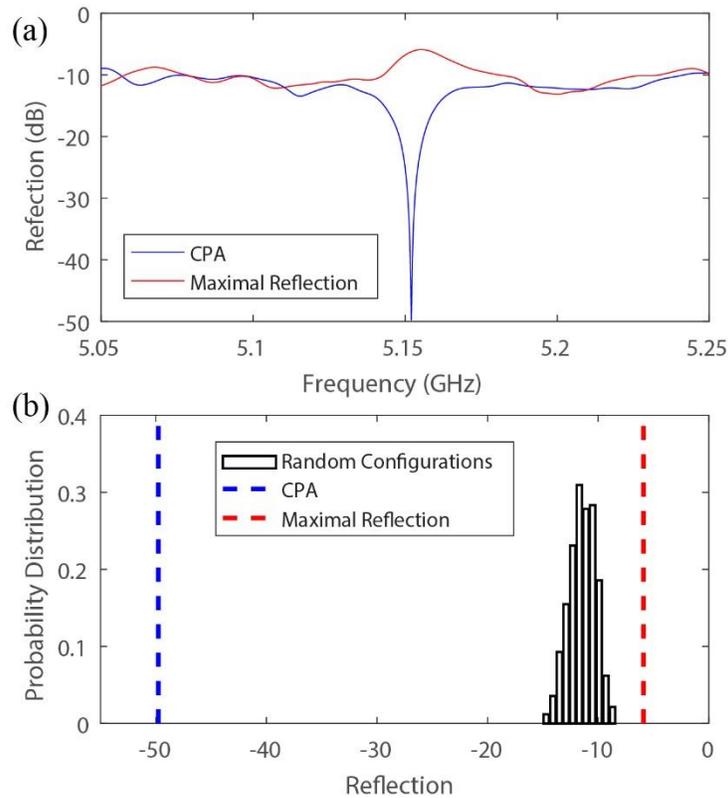

**Fig. 4.** (a) Spectrum corresponding to CPA (blue) and anti-CPA condition (red) for frequencies within the 5.135 – 5.159 GHz range. (b) Comparison of these two scattering anomalies with the distribution of $R'$ in our system (black).

**Acknowledgements**

The authors acknowledge funding from the French "Agence Nationale de la Recherche" under reference ANR-17-ASTR-0017, from the European Union through the European Regional Development Fund (ERDF), and from the French region of Brittany and Rennes Métropole through the CPER Project SOPHIE/STIC & Ondes. M.D. acknowledges the Institut Universitaire de France. The metasurface prototypes were purchased from Greenerwave. The authors acknowledge P. E. Davy for the 3D rendering of the experimental setup in Fig. 2(a).

Control of Absorption. *Science* **2011**, *331* (6019), 889–892. https://doi.org/10.1126/science.1200735.

# SUPPLEMENTAL MATERIAL:

# Coherent wave control in complex media with arbitrary wavefronts

Philipp del Hougne, K. Brahima Yeo, Philippe Besnier, and Matthieu Davy

*Univ Rennes, INSA Rennes, CNRS, IETR - UMR 6164, F-35000, Rennes, France*

Correspondence to philipp.delhougne@gmail.com or matthieu.davy@univ-rennes1.fr.

I. Experimental Details

    A. *Design of Programmable Meta-Atoms.*
    B. *Experimental Setup.*
    C. *Optimal Metasurface Configuration.*

II. Link between WS and GWS Operator

III. Optimality of GWS-Based Focusing

IV. Anti-Focusing with the GWS Operator

V. References

# I. Experimental Details

### *A. Design of Programmable Meta-Atoms.*

For our present work, it is important that the electromagnetic response of the two meta-atom states is distinct and has a notable impact on the field; the specific details of how this is achieved are secondary. The literature refers to arrays of reflection-programmable meta-atoms as "tunable impedance surface"[1], "programmable metasurface"[2], "spatial microwave modulator"[3] or "reconfigurable intelligent surface". Our prototype consists of 304 1-bit reflection-programmable meta-atoms: a bias voltage allows us to individually configure each meta-atom to one of its two possible digitalized states, which are characterized by opposite electromagnetic responses. Specifically, at the working frequency of 5.147 GHz, the phase of the meta-atom's reflection coefficient differs by roughly $\pi$ (see Fig. S1).

The working principle follows the one outlined in Ref.[4]: each meta-atom consists of two resonators that hybridize. The resonance of one of the two resonators depends on how a p-i-n diode is biased such that the overall response of the meta-atom can be configured to be either on or off resonance at the working frequency. While the meta-atoms in Ref.[4] only offered phase-binary control over one polarization of the electromagnetic field, the prototype we use offers independent control over two orthogonal field polarizations, effectively doubling the number of programmable meta-atoms (see inset in Fig. 2(a) of the main text).

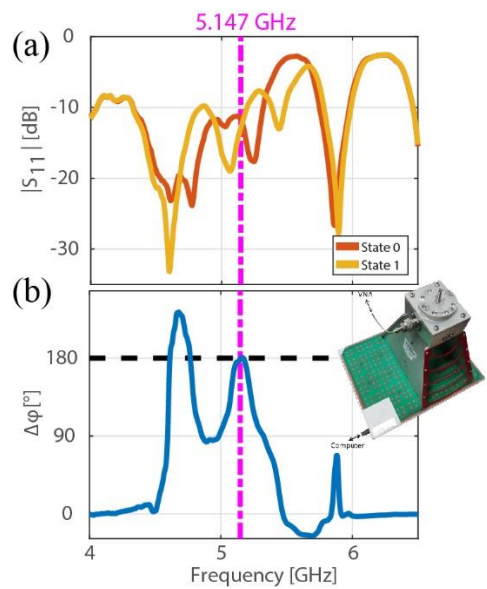

**Fig. S1**: Electromagnetic response of programmable meta-atom states "0" and "1", characterized collectively as shown in the inset, in terms of amplitude and phase of the reflected wave. At the operating frequency of 5.147 GHz, both responses have the same amplitudes (a) but their phases differ by 180° (b).

2/7

## B. Experimental Setup.

Our complex scattering enclosure is a chaotic cavity of dimensions $50 \times 50 \times 30$ cm². Two hemispheres deform the parallelepiped geometry to introduce wave chaos. If two rays are launched from the same location in slightly different directions, their separation increases exponentially with time – hence the term "wave chaotic". Based on the average decay rate of inverse Fourier transformed transmission spectra, we estimate our cavity's quality factor as $Q = 446$. This corresponds to a spectral field-field correlation length of $f_0/Q = 12$ MHz at the operating frequency $f_0 = 5.147$ GHz. A total of eight ports (coaxial connectors that are well adapted at the working frequency) is connected to the system. Their spatial location happens to be regular (four ports on the left side and four ports on the right side, see Figure 2a of the main text) but is irrelevant for the scattering matrix formalism. The ports are connected to an eight-port vector network analyzer (VNA) which acquires the complete scattering matrix in one go. For the GWS focusing experiment, we connect the eighth port via an electro-mechanical switch either to an open-circuit termination or to the VNA (matched load condition). In this case, we are thus measuring a $7 \times 7$ scattering matrix and can simultaneously monitor the intensity focused on the impedance-tunable eighth port.

The scattering matrix in our experiments is strongly non-unitary because of strong absorption on the walls of the cavity. This includes losses due to the programmable metasurfaces. We measure an average reflection coefficient $R = \langle \|S\psi_{in}\|^2 \rangle = 0.045$ which is hence well below unity. Here, the averaging is performed over random incoming wavefronts $\psi_{in}$ and random configurations of the metasurfaces.

## C. Optimal Metasurface Configuration.

Given that no forward model linking the meta-atom configurations to the scattering matrix exists, we use an iterative optimization procedure similar to the one in Refs.[5,6]. We begin by evaluating the cost function $\mathcal{C}$ for 100 random configurations. The one yielding the best $\mathcal{C}$ is the starting point for our iterative optimization. At each iteration, we randomly select $z$ meta-atoms and flip their state. We then measure the new scattering matrix and evaluate the corresponding cost function. If the latter gives an improved value, the modification of the meta-atom configurations is kept. The value of $z$ is decreased over the course of the iterations according to $z = \max(\text{int}(50e^{-0.02k}),1)$, where $k$ is the iteration index. The number of required iterations depends on the amount of reverberation inside the cavity: the more reverberation there is, the more the optimal configuration of one meta-atom depends on how the other meta-atoms are configured. As usual for inverse-design problems, without having explored the entire parameter space of $2^n$ configurations ($n$ being the number of one-bit programmable meta-atoms), our algorithm rapidly identifies a local optimum. Of course, we cannot guarantee that there is not a better configuration, but typically, we find that local optima from different optimization runs are of comparable quality.

## II. Link between WS and GWS Operator

For the sake of (historic) completeness, in this section we briefly retrace how the GWS operator emerged. Initially, the time-delay operator was studied by Wigner[7] and Smith[8] in nuclear scattering. It generalizes the delay time of waves in 1D systems to multi-channel systems and is defined as $Q = -iS^{-1}(\omega)\frac{\partial S(\omega)}{\partial \omega}$. An eigenstate of this operator is referred to as time-delay eigenstate because upon injecting this wavefront into the system a specific time delay given by the real value of the associated eigenvalue will be observed. More recently, this operator was also applied to mesoscopic physics[9,10]. Besides the creation of particle-like scattering states[11–13], this operator also enables completely non-invasive optimal focusing on a resonator embedded in a complex scattering medium[14,15]. The GWS operator considered in our main text generalizes the time-delay operator in that it replaces $\omega$ with an arbitrary global or local parameter $\alpha$ of the system[16]: $Q_\alpha = -iS^{-1}(\alpha)\frac{\partial S(\alpha)}{\partial \alpha}$.

## III. Optimality of GWS-Based Focusing

In the main text, we identified an analogy between our use of the targeted port's impedance as variable $\alpha$ and the use of the value of the dielectric constant of the target as variable $\alpha$. The latter was shown in Ref.[17] to enable optimal focusing on the target. In this section, we offer an alternative proof of this optimality for our GWS experiment. In principle, the phase conjugate of the $7 \times 1$ vector $t$ containing the transmission coefficients between the seven controlled ports and the target is the optimal wavefront. However, without field measurements at the target, $t$ is of course not known. Given the size of the port (half a wavelength) and its location on the wall of the cavity (no shadow effect), the perturbation $\Delta S$ can be expressed as $\Delta S = t\sigma t^T$, where $\sigma$ is a complex parameter related to the scattering strength of the port. $\Delta S$ therefore constitutes a rank-one perturbation of $S$ so that $Q_\alpha$ is also a rank-one matrix: $Q_\alpha = [-iS_{OC}^{-1}t\sigma]t^T$. The left eigenvector of $Q_\alpha$ verifying $q_1^\dagger Q = \tau_1 q_1^\dagger$ with eigenvalue $\tau_1 = [-i\sigma t^T S_{OC}^{-1} t]$ is therefore identified as the phase-conjugate of the transmission vector $t$, $q_1 = t^*/\|t\| = \psi_{pc}$. The GWS operator hence identifies *non-invasively* the optimal wavefront for focusing on the target.

## IV. Anti-Focusing with the GWS Operator

We also used the approach based on the GWS operator to minimize the intensity (anti-focusing) on the targeted port. In this case, the transmitted intensity $|t^T \psi_{\text{arb}}|^2$ is expected to vanish if the incoming wavefront is orthogonal to $q_1$. We hence extract the anti-focusing wavefront by minimizing the correlation between $q_1$ and $\psi_{\text{arb}}$. Experimentally, we obtain an average anti-focusing intensity of $9.6 \times 10^{-7}$ which corresponds to an average reduction of the transmission to the target by a factor of 735 compared to unoptimized systems.

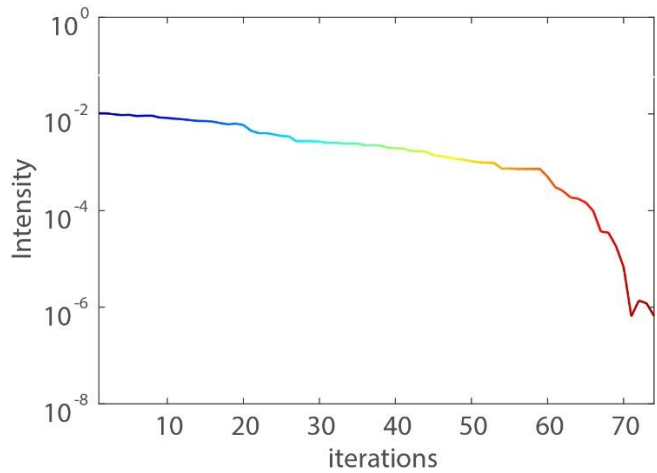

**Fig. S2**: Example of an iterative optimization which minimizes $\mathcal{C}_{GWS} = |q'_1 \psi_{\text{arb}}^\dagger|$. The intensity on the target is seen to decreases over the course of the optimization process reaching $9.6 \times 10^{-7}$.